\documentclass[proceedings]{crckbked}    

\usepackage{amsfonts}
\usepackage{amsmath}
\usepackage{graphicx}
\usepackage{courier}
\usepackage[symbol]{footmisc}

\renewcommand\arraystretch{1}
\DeclareMathAlphabet{\mathsf}{OT1}{cmss}{bx}{n}
\DeclareMathAlphabet{\mathit}{OT1}{cmr}{bx}{it}
\def\limfunc#1{\mathop{\rm #1}}%
\def\unit#1{\mathop{\rm #1}\nolimits}%
\makeatletter%
\@ifundefined{mathletters}{%
  \newcounter{equationnumber}
  \def\mathletters{%
     \addtocounter{equation}{1}
     \edef\@currentlabel{\theequation}%
     \setcounter{equationnumber}{\c@equation}
     \setcounter{equation}{0}%
     \edef\theequation{\@currentlabel\noexpand\alph{equation}}%
  }
  
}{}%
\makeatother

\begin{document}
\begin{opening}

\title{Constrained-Hamiltonian Shallow-Water \\
Dynamics on the Sphere\thanks{To appear \textit{in} O. U.
Velasco-Fuentes et al. (eds.), \textit{Nonlinear Processes in
Geophysical Fluid Dynamics}, Kluwer
Academic.}}%
\author{F.~J. \surname{Beron-Vera}}%
\runningauthor{F.~J. Beron-Vera}%
\runningtitle{Constrained Dynamics on the Sphere}%
\institute{RSMAS, University of Miami, Florida, USA}%

\begin{abstract}
Salmon's nearly geostrophic model for rotating shallow-water flow
is derived in full spherical geometry. The model, which results
upon constraining the velocity field to the height field in
Hamilton's principle for rotating shallow-water dynamics,
constitutes an important prototype of Hamiltonian balanced models.
Instead of Salmon's original approach, which consists in taking
variations of particle paths at fixed Lagrangian labels and time,
Holm's approach is considered here, namely variations are taken on
Lagrangian particle labels at fixed Eulerian positions and time.
Unlike the classical quasigeostrophic model, Salmon's is found to
be sensitive to the differences between geographic and geodesic
coordinates. One consequence of this result is that the $\beta $
plane approximation, which is included in Salmon's original
derivation, is not consistent for this class of model.
\end{abstract}


\end{opening}

\section{Introduction}

The rotating shallow-water (SW) equations constitute a pa\-radigm
for geophysical fluid motions ranging from fast timescale
dynamics, associated with inertia-gravity waves, to slow
advective-timescale dynamics, associated with nonlinear vortical
motions and Rossby waves (cf. \opencite{Gill-82};
\opencite{Pedlosky-87}). This set of equations constitute the\
``primitive''\ equations on which different approximations are
usually performed. In this paper I deal with those approximations
which involve the introduction of balance relations or constraints
that lead to filtering out the fast degrees of freedom.\ Terms
commonly used to denote the resulting models are ``balanced,''
``constrained,'' or ``intermediate;'' the latter, in particular,
reflects the fact of being at a level which is in between the
primitive equations and the equations for geostrophic motion. For
an extensive review on the wide variety of balanced models that
exists in the literature the reader is referred to
\inlinecite{Allen-Barth-Newberger-90a}.

Of particular interest are those balanced models derived by
performing approximations directly in Hamilton's principle (HP)
for SW dynamics as proposed by Salmon(1983, hereafter referred to
as S83)\nocite{Salmon-83}. This procedure allows the fundamental
symmetry-based conservation laws of the underlying primitive
system to be preserved. The approach consists in substituting
leading order balance relations and asymptotic expansions into HP
before taking variations. In particular, S83's model is derived by
constraining the velocity field to the height field in the form of
a geostrophic balance relation, i.e. between the pressure gradient
and the Coriolis force. This so-called L1 model, however, was
shown to produce less accurate solutions to the SW equations than
those produced
by other non-Hamiltonian intermediate models \cite%
{Allen-Barth-Newberger-90a,Allen-Barth-Newberger-90b,Barth-Allen-Newberger-90}%
. This is indicative of the known fact that possession of Hamiltonian
structure is no guarantee of model's accuracy. Nevertheless, other balance
relation choices---potentially more accurate than that considered by
S83---are possible \cite{Allen-Holm-96,Allen-Holm-Newberger-02}. This fact
makes the L1 model an important prototype of constrained Hamiltonian models,
and thus motivates the present study.

The L$1$ nearly geostrophic model, as well as its relatives the
extended-geostrophic Hamiltonian models of
\inlinecite{Allen-Holm-96} and
\inlinecite{Allen-Holm-Newberger-02}, have been derived in the
Cartesian coordinates of the $\beta $ plane approximation. Such
approximation relies upon expansion of the equations of motion
with respect to geographic (e.g. spherical longitude and latitude)
coordinates about some fixed point on the surface of the planet,
in inverse powers of the (mean) radius of the planet. The
expansion is then truncated at first order but retaining only the
first order variation of the Coriolis parameter (the so called
$\beta $ term) and neglecting all metric terms, which are of the
same order as the $\beta $ term! Consequently, the $\beta $ plane
approximation is only valid locally and in geodesic coordinates
\cite{Phillips-73,Verkley-90b}. These coordinate systems are such
that all the derivatives of the metric tensor vanishes identically
at the origin and thus locally look like Cartesian coordinates.
Geographic coordinates are not geodesic in general, except at the
equator where coordinate curves are geodesic curves, e.g. great
circles in spherical geometry. Consequently, only at the equator
the $\beta $ plane approximation is valid when written in
geographic coordinates, but this region is forbidden for the L1
model.

Remarkable is the fact that the quasigeostrophic (QG)
model---perhaps the most exploited model of (slow
advective-timescale) intermediate dyna- mics---has the property of
being insensitive to differences between geographic and geodesic
coordinates, namely the $\beta $ plane approximation gives the
right QG equations (\opencite{Pedlosky-87};
\opencite{Ripa-JPO-97b}, hereafter referred to as R97). Even
though the QG system does not fit within the frame of models of
the L1 class, i.e. it does not follow from an approximation made
in HP for SW motion, it can be derived from HP but for stationary
variations of a particularly chosen action
\cite{Virasoro-81,Holm-Zeitlin-98}.

The goal of this paper is to derive an L$1$ model using
non-Cartesian geometry in order to make an assessment of the
sensitivity of this model to the difference between geographic and
geodesic coordinates. I am not aware of a similar development
within the Hamiltonian framework except for the works of
\inlinecite{Shutts-89} and \inlinecite{Verkley-01}.
\inlinecite{Shutts-89} derived a modified version of the Hoskins'
(1975)\nocite{Hoskins-75} semigeostrophic equations, which are
another type of intermediate equations that can be derived from
the L1 model through a transformation into ``geostrophic
coordinates'' \cite{Salmon-85}. \inlinecite{Verkley-01}, in turn,
presented a derivation of an isentropic L1-type model for
application to atmospheric flows; the model derived here is based
on SW dynamics. Unlike both \citeauthor{Shutts-89}'
\shortcite{Shutts-89} and \citeauthor{Verkley-01}'s
\shortcite{Verkley-01} derivations, in this paper I use tools from
non-Cartesian tensor algebra, which leads to an invariant
formulation for the dynamical equations of the L1 model.

The reminder of the paper is organized as follows. In \S\,2 I set
up a mathematical model for the Earth's surface that defines the
space in which the analysis is carried out. Section 3 includes a
derivation of the general equations for a free particle on the
smooth surface of the Earth in invariant form. This is done from
HP for a general spheroidal Earth in \S\,3.1. The usual spherical
approximation is then applied to the resulting motion equations,
which, in particular, are written in geographic coordinates
(\S\,3.2). Section 3.3 presents a discussion of the consistency of
the so-called planar approximations, which include the classical
$f$ and $\beta $. Section 4 is devoted to extending
into non-Cartesian geometry Holm's (1996, hereafter referred to as H96)%
\nocite{Holm-96} general HP for variations of Lagrangian particle
labels at fixed Eulerian positions and time. The SW and L1 model
equations are derived in \S\S\,4.4 and 4.5, respectively, using
the spherical Earth's model. The equations are written in a
coordinate-invariant fashion on the sphere and then particularized
to the common geographic coordinate system. Concluding remarks are
given in \S\,5. Appendix A presents various relationships involved
in the derivation of the equations. Appendix B is reserved for the
discussion and comparison of alternative HPs.

\section{Earth's Shape Model\label{EarthModel}}

\setcounter{equation}{0}
\renewcommand{\theequation}{\thesection.\arabic{equation}}

I consider here some basic geophysical facts that relate to the
shape of the Earth and the forces acting on its equilibrium
surface (e.g. \opencite{Stommel-Moore-89}; \opencite{Ripa-RMF-95};
\opencite{Ripa-FCE-96}; R97). The mathematical framework on which
the invariant formulation of the equations derived in this paper
is based involve concepts from non-Cartesian tensor algebra (e.g.
\opencite{Abraham-Marsden-Ratiu-88};
\opencite{Dubrovin-Fomenko-Novikov-92a}) that I start by reviewing
first.

\subsection{Non-Cartesian Tensor Algebra Background\label{background}}

Let $S$ be a two-dimensional manifold, coordinatized by $\mathbf{x}%
:=(x^{1},x^{2})$. Two-dimensional intrinsic vectors on $S$ at any point $%
\mathbf{x}$ define the\textit{\ }tangent space, $T_{\mathbf{x}}S$. The
disjoint union of tangent spaces constitute the tangent bundle, $TS$. Let $%
\{e_{i}\}$ be a basis for $T_{\mathbf{x}}S$ and $\{e^{i}\}$ for the\textit{\
}dual space, $(T_{\mathbf{x}}S)^{\ast }$, namely
\begin{equation}
e^{i}(e_{j})=\delta _{j}^{i},
\end{equation}%
where $\delta _{j}^{i}$ are the Kroenecker symbols which equal 1 if $i=j$
and $0$ otherwise. Let $T_{n}^{m}(T_{\mathbf{x}}S)$ be the space of $m$%
-contravariant and $n$-covariant real valued tensors or, simply, $(m,n)$%
-tensors. Vectors $a\in T_{0}^{1}(T_{\mathbf{x}}S)=T_{\mathbf{x}}S$ are
expressed as $a=a^{i}e_{i}$ and covectors $\alpha $\textit{\textbf{%
\thinspace }}$\in T_{1}^{0}(T_{\mathbf{x}}S)=(T_{\mathbf{x}}S)^{\ast }$ as $%
\alpha =\alpha _{i}e^{i};$ the quantities $a^{i}=a(e^{i})$ and $\alpha
_{i}=\alpha (e_{i})$ are the components of $a$ and $\alpha ,$ respectively.
(N.B. The convention of summation over repeated lower and upper indices is
understood.) In general, a $(m,n)$-tensor $A$ expresses as%
\begin{equation}
A=A_{j_{1}\cdots j_{n}}^{i_{1}\cdots i_{m}}e_{i_{1}}\otimes \cdots \otimes
e_{i_{m}}\otimes e^{j_{1}}\otimes \cdots e^{j_{n}},
\end{equation}%
where $A_{j_{1}\cdots j_{n}}^{i_{1}\cdots i_{m}}=A(e^{i_{1}},\cdots
,e^{i_{m}},e_{j_{1}},\cdots ,e_{j_{n}})$ and $\otimes $ denotes the tensor
product.

Assume now that $S$ is endowed with a Riemannian metric, namely a symmetric,
positive definite, bilinear form%
\begin{equation}
\langle \hspace{-0.02in}\langle \cdot ,\cdot \rangle \hspace{-0.02in}\rangle
:=m_{ij}\,e^{i}\otimes e^{j},
\end{equation}%
where $m_{ij}(\mathbf{x}\mathbf{)}:=\langle \hspace{-0.02in}\langle
e_{i},e_{j}\rangle \hspace{-0.02in}\rangle $. The inner product of two
vectors $a,b\in T_{\mathbf{x}}S$ is computed with respect to the metric,
i.e.
\begin{equation}
\langle \hspace{-0.02in}\langle a,b\rangle \hspace{-0.02in}\rangle =\left(
m_{ij}\,e^{i}\otimes e^{j}\right) \cdot
(a,b)=m_{ij}\,e^{i}(a)\,e^{j}(b)=m_{ij}\,a^{i}b^{j}.
\end{equation}%
In particular, the square of the distance between two nearby positions on $S$%
, $\mathbf{x}$ and $\mathbf{x}+\mathrm{d}\mathbf{x},$ is given by
\begin{equation}
\mathrm{d}s^{2}=\langle \hspace{-0.02in}\langle \mathrm{d}\mathbf{x},\mathrm{%
d}\mathbf{x}\rangle \hspace{-0.02in}\rangle =\left\| \mathrm{d}\mathbf{x}%
\right\| ^{2}=m_{ij}\,\mathrm{d}x^{i}\mathrm{d}x^{j}.
\end{equation}

Let $^{\flat }$ be the index lowering operator, and $^{\natural }$, its
inverse, be the index raising operator, which are defined by
\begin{equation}
^{\flat }:T_{\mathbf{x}}S\rightarrow (T_{\mathbf{x}}S)^{\ast };\text{\ }%
a\mapsto \langle \hspace{-0.02in}\langle a,\cdot \rangle \hspace{-0.02in}%
\rangle \quad \text{and}\quad ^{\natural }:(T_{\mathbf{x}}S)^{\ast
}\rightarrow T_{\mathbf{x}}S,
\end{equation}%
respectively. The matrix of $^{\flat }$ is $[m_{ij}],$ i.e. $(a^{\flat
})_{i}=m_{ij}\,a^{j}=:a_{i},$ whereas that of $^{\natural }$ is $%
[m_{ij}]^{-1}=m^{-1}\limfunc{adj}[m_{ij}]=:[m^{ij}]$, i.e. $(\alpha
^{^{\natural }})^{i}=m^{ij}\alpha _{j}=:\alpha ^{i}.$ Here, $m:=\det
[m_{ij}] $ and $\limfunc{adj}$ denotes adjoint (transpose cofactor).

Let, in addition, reserve the symbol $\mathbf{d}$ to denote the exterior
derivative (or generalized gradient operator), whose action on a
skew-symmetric $(0,k)$-tensor or $k$-form $\alpha ,$ i.e.%
\begin{equation}
\alpha =\alpha _{i_{1}\cdots i_{k}}\,e^{i_{1}}\wedge \cdots \wedge e^{i_{k}},
\end{equation}%
where $\wedge $ denotes the exterior product, is defined by%
\begin{equation}
\mathbf{d}\alpha :=\sum_{j,i_{1}<\cdots <i_{k}}\partial _{j}\alpha
_{i_{1}\cdots i_{k}}\,e^{j}\wedge e^{i_{1}}\wedge \cdots \wedge
e^{i_{k}}.
\end{equation}%
Notice that, in particular, if $k=0$ then $\alpha $ is simply a scalar and,
hence, $\mathbf{d}\alpha =\alpha _{,i}e^{i}=:\limfunc{grad}\alpha .$ N.B.
The shorthand notations $\partial _{i}(\cdot )$ and $(\cdot )_{,i}$ for
partial differentiation $\partial (\cdot )/\partial x^{i}$ are in use.

Finally, let $\mathbb{P}$ be a linear map, with matrix elements $\mathbb{P}%
_{ij}=\sqrt{m_{ij}}$ for $i=j$ and $\mathbb{P}_{ij}=0$ otherwise. Then $%
\mathbb{P}\cdot a\mathbf{\ }$(resp., $\mathbb{P}^{-1}\cdot a^{\flat }$)
denotes the physical---nontensorial---contravariant (resp., covariant)
counterpart of vector $a$. For orthogonal coordinates, i.e. with $m_{ij}=0$
for $i\neq j$, physical contravariant and covariant counterparts coincide,
namely $\mathbb{P}\cdot a\equiv \mathbb{P}^{-1}\cdot a^{\flat }$.

\subsection{General Assumptions on $S$}

Two main assumptions make the two-dimensional manifold $S$ an
idealized model of the surface of the (solid) Earth. First, $S$ is
assumed to be embedded in a three-dimensional Euclidean space
which rotates steadily, with spinning frequency $\Omega ,$ with
respect to a Newtonian inertial space. Second, $S$ is assumed to
be a geopotential surface. Namely the projections onto
$T_{\mathbf{x}}S$ of the\textit{\ \textbf{centrifugal force} }(due
to the spinning of the planet with respect to an inertial
reference frame) and the\textit{\ \textbf{gravitational
attraction} }(due to the deviation of the shape of the planet from
a perfect sphere and to inhomogeinities in the mass distribution
within the planet) are assumed to balance one another exactly.

As a consequence of the second assumption it follows that%
\begin{equation}
\Phi :=V+V_{\mathrm{C}}=\mathrm{const.}
\end{equation}%
on $S,$ where $V$ is the\textit{\ \textbf{gravitational potential}}, $V_{%
\mathrm{C}}$ stands for the\textit{\ \textbf{centrifugal potential}}, and
their sum, $\Phi ,$ defines the\textit{\ \textbf{geopotential}.} (The
constant in the above expression is arbitrary and can be freely set to
zero.) The centrifugal potential (per unit mass) can be expressed in
invariant form as%
\begin{equation}
V_{\mathrm{C}}=-\tfrac{1}{2}\left\| \sigma \right\| ^{2},
\end{equation}%
where $\sigma $ is the velocity of the $\mathbf{x}$-system with respect to a
suitable inertial frame.

Finally, the\textit{\ \textbf{acceleration of gravity }}is defined as the
minus gradient of $\Phi $, thereby determining the vertical direction at
each point $\mathbf{x}$ on $S$. Its magnitude is thus given by
\begin{equation}
g(\mathbf{x})=\left\| \limfunc{grad}\Phi \right\| =\sqrt{m^{ij}\Phi
_{,i}\Phi _{,j}}.
\end{equation}

\subsection{Spherical Model}

It is convenient---and quite accurate---to consider $S$ as a
(two-dimensional) sphere of radius $R$, say, but\textit{\ }keeping\textit{\ }%
the main effect of the gravitational force. Namely that it can sustain a
steady rotation, relative to an inertial frame, in any point on $S$. Thus
let the coordinates on $S$ be given by
\begin{equation}
x^{1}=\left( \lambda -\lambda _{0}\right) R\cos \vartheta _{0},\quad
x^{2}=\left( \vartheta -\vartheta _{0}\right) R,  \label{xy}
\end{equation}
which are rescaled longitude, $\lambda $, and latitude, $\vartheta $, that
will be referred here to as \textit{\textbf{geographic coordinates}}. In
this case one can introduce the usual notations $(x,y)$ for $(x^{1},x^{2})$
and $(\mathbf{\hat{x}},\mathbf{\hat{y}})$ for $(e^{1},e^{2}).$ The
corresponding metric matrix, velocity of the $\mathbf{x}$-system, and
centrifugal potential, respectively, read:
\begin{equation}
\lbrack m_{ij}]=\left[
\begin{array}{cc}
\gamma ^{2} & 0 \\
0 & 1%
\end{array}%
\right] ,\quad \sigma =\frac{f}{2\tau \gamma }\mathbf{\hat{x}},\quad V_{%
\mathrm{C}}=-\frac{f^{2}}{8\tau ^{2}}.
\end{equation}%
Here,

\begin{equation}
f:=2\Omega \sin \vartheta ,\quad \gamma :=\sec \vartheta _{0}\cos \vartheta
,\quad \tau :=R^{-1}\tan \vartheta ;
\end{equation}%
the first parameter is the Coriolis parameter whereas the other two are the
geometric coefficients as defined by Ripa (2000a,b)\nocite%
{Ripa-JPO-00a,Ripa-JPO-00b}. Consistently with this spherical
approximation, the acceleration of gravity is taken as a constant,
namely $g\approx 9.8$ $\unit{m}^{2}\unit{s}^{-1}.$

More accurate models (not treated here) should account explicitly for the
flattening of the planet at the poles. For instance, although still crude,
next in accuracy can be mentioned one that has the form of an axisymmetric
spheroid of revolution (Chandrasekhar, 1969; cf. also R97)\nocite%
{Chandrasekhar-69}.

\section{Particle Dynamics\label{PartDyn}}

\setcounter{equation}{0}
\renewcommand{\theequation}{\thesection.\arabic{equation}}

In this section the manifold $S$ is assumed to represent a smooth
and frictionless Earth's surface on which a particle moves freely.
The derivation of the particle's equations of motion is
instructive inasmuch as it sets the grounds for tackling the more
complicated problem of the following section. In particular, it
shows clearly how the Coriolis force---which finds its origin in
the gravitational force---arises directly from a HP with an action
appropriate for an inertial observer, but written in coordinates
fixed to the planet. The method is in essence the same as the one
used by Pierre Simon de Laplace (1749--1827) to introduce this
force over quarter a century before than Gaspard Gustave de
Coriolis (1792--1843) was born (cf. R95; R96). The analysis of the
particle's equations allows, in addition, one to simplify the
discussion on the consistency of the so-called planar
approximations (cf. R97).

\subsection{General Equations\label{Spheroid}}

From an inertial observer viewpoint, the only force acting on the particle
is the gravitational one. The particle's kinetic and potential energies (per
unit mass) as measured by this observer are given by
\begin{equation}
T(\mathbf{x},\mathbf{\dot{x}}):=\tfrac{1}{2}\left\| \mathbf{\dot{x}}+\sigma
\right\| ^{2},\quad V(\mathbf{x})=-V_{\mathrm{C}}=\tfrac{1}{2}\left\| \sigma
\right\| ^{2},
\end{equation}%
respectively, where the overdot denotes time differentiation and a zero
value of the geopotential has been assigned to the Earth's surface. The
Lagrangian function, $L:TS\rightarrow \mathbb{R},$ is constructed in the
usual way, i.e.
\begin{equation}
L(\mathbf{x},\mathbf{\dot{x}}):=T-V=\tfrac{1}{2}\left\| \mathbf{\dot{x}}%
\right\| ^{2}+\langle \hspace{-0.02in}\langle \mathbf{\dot{x}},\sigma
\rangle \hspace{-0.02in}\rangle .  \label{L-Particle}
\end{equation}%
Let $\delta t$ be a time displacement and $\delta \mathbf{x}:=\left. \mathrm{%
d}/\mathrm{d}\varepsilon \right| _{\varepsilon =0}\mathbf{x}(t+\varepsilon
\delta t)$ a variation of the curve $\mathbf{x}:[t_{0},t_{1}]\rightarrow T_{%
\mathbf{x}}S.$ Let, in addition,
\begin{equation}
\mathcal{S}[\mathbf{x}\mathbf{]}:=\int_{t_{0}}^{t_{1}}\mathrm{d}t\,L:%
\mathcal{F}([t_{0},t_{1}])\rightarrow \mathbb{R}
\end{equation}%
be the action functional, where $\mathcal{F}([t_{0},t_{1}])$ denotes the set
of sufficiently smooth real valued functions on$\mathcal{\ }[t_{0},t_{1}]$.
Subject to fixed endpoint conditions, i.e. $\delta \mathbf{x}%
(t_{0})=0=\delta \mathbf{x}(t_{1})$, the first variation of $\mathcal{S},$
defined as $\delta \mathcal{S}:=\left. \mathrm{d}/\mathrm{d}\varepsilon
\right| _{\varepsilon =0}\mathcal{S}[\mathbf{x}+\varepsilon \delta \mathbf{x}%
],$ is given by
\begin{eqnarray}
\hspace{-1in}\delta \mathcal{S}\hspace{-0.075in}&=&\hspace{-0.075in}\int_{t_{0}}^{t_{1}}%
\hspace{-0.05in}\mathrm{d}t\,\left( L_{,i}\delta x^{i}+\frac{\partial L}{\partial \dot{x}^{i}%
}\delta \dot{x}^{i}\right) \notag \\
\hspace{-0.075in}&=&\hspace{-0.075in}\int_{t_{0}}^{t_{1}}\hspace{-0.05in}\mathrm{d}t\,\left( L_{,i}-%
\frac{\mathrm{d}}{\mathrm{d}t}\frac{\partial L}{\partial
\dot{x}^{i}}\right)
\delta x^{i}  \notag \\
\hspace{-0.075in}
&=&\hspace{-0.075in}\int_{t_{0}}^{t_{1}}\hspace{-0.05in}\mathrm{d}t\,\left[
\tfrac{1}{2}\left( m_{jk,i}-m_{ki,j}-m_{ji,k}\right) \dot{x}^{j}\dot{x}^{k}-%
\ddot{x}_{i}-\left( \sigma _{i,j}-\sigma _{j,i}\right)
\dot{x}^{j}\right] \hspace{-0.025in}\delta x^{i}.
\end{eqnarray}%
HP ($\delta \mathcal{S}=0$) then yields the Newton's law for the
particle in covariant form%
\begin{equation}
\fbox{$\mathrm{D}\mathbf{\dot{x}}^{\flat }/\mathrm{d}t+\mathbf{d}\sigma
^{\flat }\cdot \mathbf{\dot{x}}=0.$}  \label{part_cov}
\end{equation}%
In this equation, the coordinate representation of the object $\mathrm{D}%
\alpha /\mathrm{d}t,$ for any covector $\alpha ,$ is given by $(\mathrm{D}%
\alpha /\mathrm{d}t)_{i}=\dot{\alpha}_{i}-\Gamma _{ij}^{k}\dot{x}^{j}\alpha
_{k}$, where $\Gamma _{ij}^{k}(\mathbf{x}):=\tfrac{1}{2}%
m^{kl}(m_{il,j}+m_{jl,i}$\allowbreak $-m_{ij,l})$ are the Christoffel
symbols (of second kind), which establish the (Levi--Civita) connection on
the Riemannian manifold $S$. In addition,%
\begin{equation}
\mathbf{d}\sigma ^{\flat }=\sigma _{i,j}\,e^{i}\wedge e^{j}=\left( \sigma
_{i,j}-\sigma _{j,i}\right) \,e^{i}\otimes e^{j},  \label{coriolis}
\end{equation}%
which can be regarded as the\textit{\ \textbf{Coriolis two-form}. }Notice
that $\mathbf{d}\sigma ^{\flat }\cdot \mathbf{\dot{x}}$ $=$ $[\left( \sigma
_{i,j}-\sigma _{j,i}\right) \,e^{i}\otimes e^{j}]\cdot \mathbf{\dot{x}}$ $=$
$\left( \sigma _{i,j}-\sigma _{j,i}\right) \,e^{i}(\cdot )\,e^{j}(\mathbf{%
\dot{x}})$ $=$ $\left( \sigma _{i,j}-\sigma _{j,i}\right) \dot{x}^{j}$%
\thinspace $e^{i}.$ The operator $^{^{\natural }}$ transforms (\ref{part_cov}%
) into its contravariant counterpart
\begin{equation}
\mathrm{D}\mathbf{\dot{x}/}{\mathrm{d}}t+(\mathbf{d}\sigma ^{\flat }\cdot
\mathbf{\dot{x}})^{^{\natural }}=0;  \label{part_con}
\end{equation}%
here, $\mathrm{D}a/\mathrm{d}t,$ for any covector $a,$ in components reads $(%
\mathrm{D}a/\mathrm{d}t)^{i}=\dot{a}^{i}+\Gamma _{jk}^{i}\dot{x}^{j}a^{k}.$

Equations (\ref{part_cov}) or (\ref{part_con}) are invariant under general
coordinate transformations on $S$, which in this case is not restricted to
the spherical Earth model. In particular, these equations nicely show that
the Coriolis term is responsible for the particle's trajectory to depart
from a geodesic curve on $S$, i.e. a pure Galilean inertial motion. The
latter is only consistent with motions with sufficiently large initial
kinetic energy as shown by R97, who described all possible solutions on a
sphere, namely the so-called inertial oscillations.

\subsection{Equations on the Sphere in Geographic Coordinates \label{Sphere}}

In the geographic coordinate system (\ref{xy}) of the spherical Earth's
model the only nonzero Christoffel symbols are $\Gamma _{11}^{2}=\gamma
^{2}\tau $ and $\Gamma _{12}^{1}=\Gamma _{21}^{1}=-\tau .$ In turn, the
matrix of the Coriolis two-form takes the form
\begin{equation}
\lbrack \sigma _{i,j}-\sigma _{j,i}]=\left[
\begin{array}{cc}
0 & -\gamma f \\
\gamma f & 0%
\end{array}%
\right] .
\end{equation}%
Thus equations (\ref{part_cov}), with the spherical approximation and
particularized to geographic coordinates, take the following component
representation:
\renewcommand{\arraystretch}{1.25}%
\begin{equation}
\left.
\begin{array}{r}
\ddot{x}_{1}-\gamma ^{2}\tau \dot{x}^{1}\dot{x}_{2}-(\gamma f+\tau \dot{x}%
_{1})\dot{x}^{2}=0, \\
\ddot{x}_{2}+(\gamma f+\tau \dot{x}_{1})\dot{x}^{1}=0.%
\end{array}%
\hspace{-0.05in}\right\}  \label{part-esf}
\end{equation}%
\renewcommand{\arraystretch}{1.00}%
Let $\mathbf{u}:=(u,v):=\mathbb{P}\cdot \mathbf{\dot{x}}$ ($\equiv \mathbb{P}%
^{-1}\cdot \mathbf{\dot{x}}^{\flat }$ since the coordinates are orthogonal;
cf. \S\,2.1). Application of $\mathbb{P}^{-1}$ transforms set (%
\ref{part-esf}) into the more familiar form (e.g. R97)%
\renewcommand{\arraystretch}{1.25}%
\begin{equation}
\left.
\begin{array}{r}
\dot{u}-(f+\tau u)v=0, \\
\dot{v}+(f+\tau u)u=0,%
\end{array}%
\hspace{-0.05in}\right\}
\end{equation}%
\renewcommand{\arraystretch}{1.00}%
which can be written in vector notation as well, i.e.
\begin{equation}
\mathbf{\dot{u}}+(f+\tau u)\,\mathbf{\hat{z}}\times \mathbf{u}=0,
\end{equation}%
where $\mathbf{\hat{z}}$ is the vertical unit vector and $\times $ denotes
the cross product of vectors.

\subsection{``Planar'' Approximations\label{Beta}}

In addition to the spherical approximation, other standard
approximations introduced in the equations are the ``planar''
approximations. These approximations, which are meant to be valid
locally at a point on the sphere in geographic coordinates, are
obtained by expanding the equations in inverse powers of the
radius of the sphere $R$. The most common approximations being the
$f$ and $\beta $. The former is a consistent zeroth-order
approximation. The latter, however, is an inconsistent first-order
approximation, except at the equator. A consistent $n$th-order
approximation is understood as one that produces $O(R^{-n-1})$
errors in the integrals of motion associated with the equations on
the sphere. These integrals are the (kinetic) energy of the
particle as measured by a terrestrial observer,%
\begin{equation}
E:=\tfrac{1}{2}\mathbf{u}^{2},
\end{equation}%
and the absolute angular momentum (with respect to the center of
the planet and in the direction of the axis of rotation), which,
up to some constants, is
given by%
\begin{equation}
M:=\gamma u-\Omega R\left( \cos \vartheta _{0}-\gamma \cos \vartheta \right)
.  \label{AngMom}
\end{equation}

R97 showed that a consistent first-order ``planar'' approximation must have
\begin{equation}
\text{\textit{\textbf{Ripa} \textbf{``plane''}}}:\gamma =1-\tau _{0}y,\;\tau
=\tau _{0}/\gamma ,\;f=f_{0}+\beta y/\gamma
\end{equation}%
where $\tau _{0}:=R^{-1}\tan \vartheta _{0},\;f_{0}:=2\Omega \sin \vartheta
_{0},$ and $\beta :=2\Omega R^{-1}\cos \vartheta _{0}$. With this
approximation the equations of motion conserve $\frac{1}{2}\mathbf{\dot{x}}%
^{2}-\tau _{0}y\dot{x}^{2}=E-O(R^{-2})$ and $(1-\tau _{0}y)u-f_{0}y-\frac{1}{%
2}\beta (1-R^{2}\tau _{0}^{2})y^{2}=M-O(R^{-2})$. The $f$ plane approximation%
\textit{\ }has
\begin{equation}
\mathit{f}\text{\textit{\ \textbf{plane}}}:\gamma =1,\text{\ }\tau
=0,\;f=f_{0},
\end{equation}%
which consistently implies conservation of $\frac{1}{2}\mathbf{\dot{x}}%
^{2}=E-O(R^{-1})$ and $u-f_{0}y=M-O(R^{-1}).$ The\textit{\ }$\beta $ plane
approximation, in turn, has
\begin{equation}
\boldsymbol{\beta }\text{\textit{\ \textbf{plane}}}:\gamma
=1,\;\tau =0,\;f=f_{0}+\beta y
\end{equation}%
and implies conservation of $\frac{1}{2}\mathbf{\dot{x}}^{2}$ and $u-f_{0}y-%
\frac{1}{2}\beta y^{2}$, which produce $O(R^{-1})$ errors to $E$ and $M,$
respectively, everywhere except at $\vartheta _{0}=0$ where these errors are
$O(R^{-2})$ because $\tau _{0}\equiv 0$.

It is thus clear that a consistent first-order approximation must include,
in general, non-Cartesian terms in order to correctly reproduce the
conservation laws of the system. (That is the reason for the quotation marks
in this section.) It is worthwhile remarking that this is no longer
necessary for motions around the equator. Geographic coordinates at the
equator are\textit{\ \textbf{geodesic coordinates} }because all the
derivatives of the metric vanish there. For this reason locally at the
equator the geometry in geographic coordinates looks like Cartesian and,
hence, the $\beta $ plane is a consistent approximation there. In general,
for any point of a space with a symmetric affine connection coordinatized by
$x^{i},$ $i=1,2,\cdots ,$ say, there exists a coordinate system $x^{\prime
i},$ $i=1,2,\cdots $, say, such that the coefficients of the connection
vanish identically. Such a system can be defined implicitly by $%
x^{i}=x^{\prime i}-\frac{1}{2}\Gamma _{jk}^{i}(0)\,x^{\prime j}x^{\prime k}$
which can be readily seen to result in $\Gamma _{jk}^{\prime i}(0)\equiv 0.$
For geographic coordinates the transformation $(x^{\prime },y^{\prime
})\mapsto (x,y)$ reads $(x,y)=(x^{\prime },y^{\prime })+\tau _{0}x^{\prime
}(y^{\prime },-\frac{1}{2}x^{\prime })$, which reduces to the identity at $%
\vartheta =\vartheta _{0}.$ Of course, the practical use of
geodesic coordinates (away from the equator) is questionable (cf.
\opencite{Phillips-73}; \opencite{Verkley-90b}).

\section{Fluid Dynamics\label{FluDyn}}

\setcounter{equation}{0}
\renewcommand{\theequation}{\thesection.\arabic{equation}}

In this section I derive from HP the equations of motion for
(inviscid, unforced) SW and L$1$ dynamics on the spherical model
for the Earth's surface. The derivation makes use of H96's
approach but extended to non-Cartesian geometry. In this approach
variations of Lagrangian particle labels are performed at fixed
Eulerian positions and time. One advantage of H96's approach is
that the equations result directly in Eulerian coordinates.

\subsection{Lagrangian and Eulerian Coordinates}

Identification of fluid particles in a SW motion requires two-dimensional
labels $\mathfrak{x}:=(\mathfrak{x}^{1},\mathfrak{x}^{2})$, say, which are
defined in certain affine (metricless) space $\mathfrak{S}$, say. Let%
\begin{equation}
\varphi \times \limfunc{id}:\mathfrak{S}\times \mathbb{R}\rightarrow S\times
\mathbb{R};\;\left( \mathfrak{x},t\right) \mapsto (\mathbf{x},t)=\left(
\varphi (\mathfrak{x},t),t\right)
\end{equation}%
be the map that relates the Lagrangian labels with the Eulerian
two-dimensional positions at time $t$, and consider its inverse:%
\begin{equation}
\varphi ^{-1}\times \limfunc{id}:S\times \mathbb{R}\rightarrow \mathfrak{S}%
\times \mathbb{R};\;\left( \mathbf{x},t\right) \mapsto (\mathfrak{x}%
,t)=(\varphi ^{-1}(\mathbf{x},t),t).\label{inverse}
\end{equation}%
Let now $J$ and $\mathfrak{J}$ be the Jacobians of these maps, respectively,
which are defined by
\begin{subequations}
\begin{eqnarray}
J\hspace{-.05in}&:\hspace{-.09in}&=\det [J_{\mathfrak{i}}^{i}],\quad J_{%
\mathfrak{i}}^{i}:=\partial x^{i}/\partial \mathfrak{x}^{\mathfrak{i}}, \\
\mathfrak{J}\hspace{-.05in}&:\hspace{-.09in}&=\det [\mathfrak{J}_{i}^{%
\mathfrak{i}}],\quad \;\mathfrak{J}_{i}^{\mathfrak{i}}:=\partial \mathfrak{x}%
^{\mathfrak{i}}/\partial x^{i}.
\end{eqnarray}

The time derivative of a Lagrangian label, following a fluid particle, is
zero by construction. Consequently, $\mathbf{\dot{x}}=\partial _{t}\varphi
+\varphi _{,\mathfrak{i}}\mathfrak{\dot{x}}^{\mathfrak{i}}\equiv \partial
_{t}\varphi .$ The latter defines the\textit{\ \textbf{Lagrangian} \textbf{or%
} \textbf{material velocity}}
\end{subequations}
\begin{equation}
\mathfrak{v}(\mathfrak{x},t):=\partial _{t}\varphi ;
\end{equation}%
the\textit{\ \textbf{Eulerian} \textbf{or} \textbf{spatial velocity}}, in
turn, is defined by
\begin{equation}
\mathbf{v}(\mathbf{x},t):=\mathfrak{v}(\mathfrak{x},t).
\end{equation}%
Finally, the time derivative of any scalar function $a(\mathbf{x},t)$ is $%
\dot{a}=(\partial _{t}+v^{i}\partial _{i})a=:\mathrm{D}a/\mathrm{D}t,$ where
$v^{i}=\mathbf{v(}e^{i}).$

\subsection{Volume Conservation}

Let $R(t)\subset S$ be a material spherical cap (made of the same fluid
particles) and let $h(\mathbf{x},t)$ be the depth of the fluid. Let, in
addition, $h_{0}(\mathfrak{x})$ be the density of Lagrangian labels in
container $R(t)$. Since $R(t)$ is material, the Lagrangian labels are
defined in certain fixed region $\mathfrak{R}\subset \mathfrak{S}$. As a
consequence of the metricless nature of $\mathfrak{S}$, the following
equality holds:%
\begin{equation}
\int_{R(t)}\mathrm{d}^{2}\mathbf{x}\,\sqrt{m}h=\int_{\mathfrak{R}}\mathrm{d}%
^{2}\mathfrak{x}\,h_{0}.  \label{Vol}
\end{equation}%
The latter implies%
\begin{equation}
\fbox{$\sqrt{m}hJ=h_{0},$}
\end{equation}%
which is the Lagrangian form of the volume conservation law. In order to
obtain the Eulerian counterpart of this law, one needs to take the time
derivative of the l.h.s. of (\ref{Vol}), i.e.%
\begin{eqnarray}
\frac{\mathrm{d}}{\mathrm{d}t}\int_{R(t)}\mathrm{d}^{2}\mathbf{x}\,\sqrt{m}%
h &=& \int_{\mathfrak{R}}\mathrm{d}^{2}\mathfrak{x}\,\left[ \frac{\mathrm{d}J}{%
\mathrm{d}t}\sqrt{m}h+J\frac{\mathrm{d}}{\mathrm{d}t}\left( \sqrt{m}h\right) %
\right] \notag \\
&=&\int_{\mathfrak{R}}\mathrm{d}^{2}\mathfrak{x}\,\sqrt{m}J\left[
\partial _{t}h+\limfunc{div}(h\mathbf{v})\right] ,
\end{eqnarray}%
where the relationships of appendix \ref{Relations} have been
used. The
conservation law follows upon setting to zero the latter result:%
\begin{equation}
\fbox{$\partial _{t}h+\limfunc{div}(h\mathbf{v})=0,$}  \label{vol}
\end{equation}%
where $\limfunc{div}(h\mathbf{v}):=\partial _{i}\left( \sqrt{m}hv^{i}\right)
/\sqrt{m}.$ Notice that in geographic coordinates $\limfunc{div}(h\mathbf{v}%
)=\gamma ^{-1}[\partial _{x}\left( hu\right) $ $+$ $\partial _{y}\left(
\gamma hv\right) ]=:\nabla \cdot (h\mathbf{u}).$

\subsection{General HP in Eulerian Coordinates}

Following H96, I consider an action functional of the form%
\begin{equation}
\mathcal{S}[\mathfrak{x}]:=\int_{t_{0}}^{t_{1}}\mathrm{d}t\,L[\mathbf{v},%
\mathfrak{J}]=\int_{t_{0}}^{t_{1}}\mathrm{d}t\int_{D}\mathrm{d}^{2}\mathbf{x}%
\,l(\mathbf{v},\mathfrak{J};\mathbf{x}),
\end{equation}%
where $D$ is a fixed region on $S$ with solid boundary $\partial D.$ Here, $%
L $ is the Lagrangian functional and, unlike H96 who adopted Cartesian
coordinates, $l/\sqrt{m}$ is the Lagrangian density. Variations of
Lagrangian particle labels at fixed Eulerian positions and time result in%
\begin{eqnarray}
\delta \mathcal{S} &=&\int \left( \frac{\delta L}{\delta v^{i}}\delta v^{i}+%
\frac{\delta L}{\delta \mathfrak{J}}\delta \mathfrak{J}\right)   \notag \\
&=&\int \mathfrak{J}J_{\mathfrak{i}}^{i}\delta \mathfrak{x}^{\mathfrak{i}}%
\left[ \frac{\mathrm{D}}{\mathrm{D}t}\left( J\frac{\delta L}{\delta v^{i}}%
\right) +J\frac{\delta L}{\delta v^{j}}\partial _{i}v^{j}-\partial _{i}\frac{%
\delta L}{\delta \mathfrak{J}}\right]   \notag \\
&&+\int \partial _{i}\left( \delta x^{i}\mathfrak{J}\frac{\delta
L}{\delta
\mathfrak{J}}-v^{i}J_{\mathfrak{i}}^{j}\delta \mathfrak{x}^{\mathfrak{i}}%
\frac{\delta L}{\delta v^{j}}\right) \notag \\
&&-\int \partial _{t}\left( J_{\mathfrak{%
i}}^{i}\delta \mathfrak{x}^{\mathfrak{i}}\frac{\delta L}{\delta v^{i}}%
\right) ,  \label{dS}
\end{eqnarray}%
where $\int (\cdot ):=\int_{t_{0}}^{t_{1}}\mathrm{d}t\,\int_{D}\mathrm{d}%
^{2}\mathbf{x\,}(\cdot )$. Derivation of (\ref{dS}) involved the
use of the
relationships of appendix \ref{Relations}. Fixed endpoint conditions, $%
\delta \mathfrak{x}\mathbf{(}\mathbf{x},t_{1}\mathbf{)}=0=\delta \mathfrak{x}%
\mathbf{(}\mathbf{x},t_{2}\mathbf{)}$, allows one to get rid of
the last integral in (\ref{dS}). Then HP implies the motion
equation
\begin{equation}
\fbox{$\partial _{t}\left( J\dfrac{\delta L}{\delta \mathbf{v}}\right)
+\pounds _{\mathbf{v}}\left( J\dfrac{\delta L}{\delta \mathbf{v}}\right) -%
\mathbf{d}\dfrac{\delta L}{\delta \mathfrak{J}}=0$}  \label{EulerLagrange}
\end{equation}%
\textit{and }the no-flow boundary condition%
\begin{equation}
\fbox{$\langle \hspace{-0.02in}\langle \mathbf{v},n\rangle \hspace{-0.02in}%
\rangle =0\quad @\quad \partial D,$}  \label{BC-SW}
\end{equation}%
where $n$ is the external normal to the boundary. In (\ref{EulerLagrange}), $%
\pounds _{a}\alpha :=\mathbf{d}\alpha \cdot a+\mathbf{d}\langle \hspace{%
-0.03in}\langle a,\alpha \rangle \hspace{-0.03in}\rangle $ is the Lie
derivative of covector $\alpha $ along vector $a$; in components $%
(\pounds _{a}\alpha )_{i}=a^{j}\alpha _{i,j}+\alpha _{j}a_{,i}^{j}.$ Result (%
\ref{BC-SW}), in turn, made use of Gauss' theorem, namely $\int_{D}\mathrm{d}%
^{2}\mathbf{x\,}\sqrt{m}\limfunc{div}a=\int_{D}\mathrm{d}^{2}\mathbf{x\,}%
\partial _{i}(\sqrt{m}a^{i})=\oint_{\partial D}\mathrm{d}s\,a^{i}n_{i}$ for
all vector $a.$

Finally, it must be mentioned that the Euler--Poincar\'{e} formalism
provides an alternative way to obtaining (\ref{EulerLagrange})--(\ref{BC-SW}%
) \cite{Holm-Marsden-Ratiu-02}.

\subsection{HP for SW Dynamics on the Sphere\label{HPforSW}}

Under the assumption that the layer of fluid is thin enough so that it does
not represent a source of gravitation, an appropriate Lagrangian density for
a HP for SW dynamics on the sphere has%
\begin{equation}
\fbox{$l(\mathbf{v},\mathfrak{J};\mathbf{x}):=h_{0}\mathfrak{J}\left( \frac{1%
}{2}\left\| \mathbf{v}\right\| ^{2}+\langle \hspace{-0.02in}\langle \mathbf{v%
},\sigma \rangle \hspace{-0.02in}\rangle \right) -\frac{1}{2}g\sqrt{m}\left(
\dfrac{h_{0}\mathfrak{J}}{\sqrt{m}}-H\right) ^{2}$}  \label{L-SW}
\end{equation}%
along with the definitions%
\begin{equation}
\fbox{$h:=h_{0}\mathfrak{J}/\sqrt{m},\quad p:=g\left( h-H\right) .$}
\label{h&p}
\end{equation}%
Here, $h_{0}$ and $g$ are both constants, and $p(\mathbf{x},t)$ is the
hydrostatic pressure, where $H(\mathbf{x})$ is the reference depth including
the possibility of an irregular topography. The choice $h_{0}=\mathrm{%
\limfunc{const}}$\textrm{.} is necessary in order for the Lagrangian density
to be independent of the Lagrangian labels. The assumption $g=\mathrm{%
\limfunc{const}}$\textrm{.}, in turn, is consistent with the spherical
approximation for the Earth's surface. The last term on the r.h.s. of (\ref%
{L-SW}), which is not present in (\ref{L-Particle}), relates to the
gravitational potential of the fluid column due to the departure of the free
surface from the resting position.

According to%
\begin{eqnarray}
\frac{\delta L}{\delta
\mathbf{v}}\hspace{-.05in}&=&\hspace{-.05in}h_{0}\mathfrak{J}\left(
\mathbf{v}+\sigma \right) ^{\flat }, \\
 \frac{\delta L}{\delta
\mathfrak{J}}\hspace{-.05in}&=&\hspace{-.05in}h_{0}\left(
\tfrac{1}{2}\left\| \mathbf{v}\right\| ^{2}+\langle
\hspace{-0.02in}\langle \mathbf{v},\sigma \rangle
\hspace{-0.02in}\rangle -p\right) ,
\end{eqnarray}%
equations (\ref{EulerLagrange}) imply the following equivalent
sets of
equations:%
\renewcommand{\arraystretch}{1.75}%
\begin{equation}
\fbox{$%
\begin{array}{l}
\text{a. }\partial _{t}\left( \mathbf{v}+\sigma \right) ^{\flat }+\pounds _{%
\mathbf{v}}\left( \mathbf{v}+\sigma \right) ^{\flat }+\mathbf{d}\left( p-%
\tfrac{1}{2}\left\| \mathbf{v}\right\| ^{2}-\langle \hspace{-0.02in}\langle
\mathbf{v},\sigma \rangle \hspace{-0.02in}\rangle \right) =0, \\
\text{b. }\partial _{t}\mathbf{v}^{\flat }+\mathbf{d}\left( \mathbf{v}%
+\sigma \right) ^{\flat }\cdot \mathbf{v}+\mathbf{d}\left( p+\tfrac{1}{2}%
\left\| \mathbf{v}\right\| ^{2}\right) =0, \\
\text{c. }(\partial _{t}+\nabla _{\mathbf{v}})\mathbf{v}^{\flat }+\mathbf{d}%
\sigma ^{\flat }\cdot \mathbf{v}+\mathbf{d}p=0.%
\end{array}%
$}  \label{SW}
\end{equation}%
\renewcommand{\arraystretch}{1.00}%
Equation (\ref{SW}b) involves the identity $\pounds _{a}\alpha =\mathbf{d}%
\alpha \cdot a+\mathbf{d}\langle \hspace{-0.02in}\langle a,\alpha \rangle
\hspace{-0.02in}\rangle ,$ particularized for $a=\mathbf{v}$ and $\alpha
=\left( \mathbf{v}+\sigma \right) ^{\flat }.$ Equation (\ref{SW}c), in turn, $%
\mathbf{d}\langle \hspace{-0.02in}\langle a,\alpha \rangle \hspace{-0.02in}%
\rangle =\nabla _{a}\alpha +\nabla _{\alpha ^{\natural }}a^{\flat }-\mathbf{d%
}\alpha \cdot a-\mathbf{d}a^{\flat }\cdot \alpha ^{\natural }$,
specialized for $a=\mathbf{v}$ and $\alpha =\mathbf{v}^{\flat }$.
Here, $\nabla _{a}\alpha $ denotes the covariant derivative of
covector $\alpha $ in the direction of vector $a$; in components
$(\nabla _{a}\alpha )_{i}=a^{k}\alpha _{i,k}$ $-$ $\Gamma
_{ik}^{j}\alpha _{j}a^{k}.$ Any set
selected from (\ref{SW}) together with the volume conservation equation (\ref%
{vol}), all subject to the no-flow boundary condition (\ref{BC-SW}),
constitute the covariant form of the SW equations on a region $D$ defined on
the sphere. These equations (or their contravariant counterpart via the
metric) are invariant under general changes of coordinates on the sphere.

The SW system conserves energy and Casimirs, namely
\begin{equation}
\mathcal{E}:=\tfrac{1}{2}\int_{D}\mathrm{d}^{2}\mathbf{x}\,\sqrt{m}\left(
h\left\| \mathbf{v}\right\| ^{2}+p^{2}/g\right) ,\quad \mathcal{C}:=\int_{D}%
\mathrm{d}^{2}\mathbf{x}\,\sqrt{m}hC(q),
\end{equation}%
for arbitrary $C(\cdot )$ and where%
\begin{equation}
qh:=\frac{1}{\sqrt{m}h_{0}}\varepsilon ^{ij}\partial _{i}\left( J\frac{%
\delta L}{\delta v^{j}}\right) =\frac{1}{\sqrt{m}}\varepsilon ^{ij}\partial
_{i}\left( v_{j}+\sigma _{j}\right)
\end{equation}%
defines the potential vorticity $q$. The latter is conserved following fluid
particles, i.e. $\mathrm{D}q/\mathrm{D}t=0,$ as readily follows upon
noticing that%
\begin{eqnarray}
\frac{\mathrm{d}}{\mathrm{d}t}\oint_{\partial D}\mathrm{d}x^{i}\,J\frac{%
\delta L}{\delta v^{i}} &=&\oint_{\partial D}\mathrm{d}x^{i}\left[ \frac{%
\mathrm{D}}{\mathrm{D}t}\left( J\frac{\delta L}{\delta v^{i}}\right) +J\frac{%
\delta L}{\delta v^{j}}\partial _{i}v^{j}\right]   \notag \\
&=&\oint_{\partial D}\mathrm{d}x^{i}\partial _{i}\frac{\delta
L}{\delta \mathfrak{J}} \notag \\
&\equiv& 0.
\end{eqnarray}

The physical counterpart of any of the equations in (\ref{SW}) follows from
application of the inverse map $\mathbb{P}^{-1}$. In geographic coordinates,
the physical counterpart of, for instance, set (\ref{SW}b), reads (e.g. R97)%
\renewcommand{\arraystretch}{1.25}%
\begin{equation}
\left.
\begin{array}{r}
\partial _{t}u-(\xi +f)v+\gamma ^{-1}\partial _{x}B=0, \\
\partial _{t}v+(\xi +f)u+\partial _{y}B=0,%
\end{array}%
\hspace{-0.05in}\right\}  \label{SW-physical}
\end{equation}%
\renewcommand{\arraystretch}{1.00}%
where $\xi :=\gamma ^{-1}\partial _{x}v-\partial _{y}u-\tau u$ and $B:=p+%
\tfrac{1}{2}(u^{2}+v^{2})$ are the relative vorticity and
Bernoulli head, respectively. System (\ref{SW-physical}) can also
be written in vector
notation, i.e.%
\begin{equation}
\partial _{t}\mathbf{u}+(\xi +f)\,\mathbf{\hat{z}}\times \mathbf{u}+\nabla
B=0,
\end{equation}%
with $\nabla a:=(\gamma ^{-1}\partial _{x}a,\partial _{y}a)$ the gradient of
any scalar function $a(\mathbf{x})$ in geographic coordinates, and where $%
\xi =\nabla \cdot \left( \mathbf{u}\times \mathbf{\hat{z}}\right) $ and $B=p+%
\frac{1}{2}\mathbf{u}^{2}$. Finally, the integrals of motion take the form%
\begin{equation}
\mathcal{E}=\tfrac{1}{2}\int_{D}\mathrm{d}^{2}\mathbf{x}\,\gamma \left( h%
\mathbf{u}^{2}+p^{2}/g\right) ,\quad \mathcal{C}=\int_{D}\mathrm{d}^{2}%
\mathbf{x}\,\gamma hC(q),
\end{equation}%
where $q=(\xi +f)/h$ satisfies%
\begin{equation}
\mathrm{D}q/\mathrm{D}t=(\partial _{t}+u\gamma ^{-1}\partial _{x}+v\partial
_{y})q=(\partial _{t}+\mathbf{u}\cdot \nabla )q=0.  \label{qSW}
\end{equation}%
If $D$ is a zonal channel and the topography has the same symmetry, i.e. $%
\partial _{x}H\equiv 0,$ then zonal momentum,%
\begin{equation}
\mathcal{M}:=\int_{D}\mathrm{d}^{2}\mathbf{x}\,\gamma h\left[ \gamma
u-\Omega R\left( \cos \vartheta _{0}-\gamma \cos \vartheta \right) \right] ,
\end{equation}%
is also an integral of motion.

\subsection{HP for L1 Dynamics on the Sphere\label{HPforL1}}

The starting point of S83's method to derive approximate models by making
approximations in the HP for SW consists is expanding the velocity field as%
\begin{equation}
\begin{array}{ccccc}
\mathbf{v} & = & \mathbf{v}_{\mathrm{G}} & + & \mathbf{v}_{\mathrm{A}}, \\
O & : & \varepsilon &  & \varepsilon ^{2}%
\end{array}
\label{v-split}
\end{equation}%
where $\varepsilon \rightarrow 0$ is an appropriate Rossby number. The
lowest-order contribution to the velocity is assumed to satisfy the
geostrophic balance and thus is a function of the height (mass) field. In
invariant form this reads
\begin{equation}
\mathbf{v}_{\mathrm{G}}(\mathfrak{J})=-(\mathbf{d}\sigma ^{\flat
})^{-1}\cdot \mathbf{d}p  \label{vG}
\end{equation}%
(at least there where $\mathbf{d}\sigma ^{\flat }$ is invertible). The
Lagrangian density for L1 dynamics on the sphere is obtained from (\ref{L-SW}%
) after replacing $\mathbf{v}$ by (\ref{v-split}), with $\mathbf{v}_{\mathrm{%
G}}$ given by (\ref{vG}), and by dropping the $O(\varepsilon ^{4})$-term $%
\frac{1}{2}\left\| \mathbf{v}_{\mathrm{A}}\right\| ^{2}$ in the first
parenthesis. Thus%
\begin{equation}
\fbox{$l_{1}(\mathbf{v},\mathfrak{J};\mathbf{x}):=h_{0}\mathfrak{J}\left(
\langle \hspace{-0.02in}\langle \mathbf{v},\mathbf{v}_{\mathrm{G}}+\sigma
\rangle \hspace{-0.02in}\rangle -\tfrac{1}{2}\left\| \mathbf{v}_{\mathrm{G}%
}\right\| ^{2}\right) -\tfrac{1}{2}g\sqrt{m}\left( \dfrac{h_{0}\mathfrak{J}}{%
\sqrt{m}}-H\right) ^{2},$}  \label{L-L1}
\end{equation}%
together with the definitions (\ref{h&p}), gives the L1 model's Lagrangian,
i.e. $L_{1}:=\int_{D}\mathrm{d}^{2}\mathbf{x\,}l_{1}$. (A notation more
consistent with my dimensional approach should in fact be $L_{3}$ for this
Lagrangian.) According to
\begin{eqnarray}
\dfrac{\delta L_{1}}{\delta \mathbf{v}}\hspace{-.05in}&=&\hspace{-.05in}h_{0}\mathfrak{J}\left( \mathbf{v}_{%
\mathrm{G}}+\sigma \right) ^{\flat }, \\
\dfrac{\delta L_{1}}{\delta
\mathfrak{J}}\hspace{-.05in}&=&\hspace{-.05in}h_{0}\left( \langle \hspace{-0.02in}\langle \mathbf{v},\mathbf{%
v}_{\mathrm{G}}+\sigma \rangle \hspace{-0.02in}\rangle
-\tfrac{1}{2}\left\| \mathbf{v}_{\mathrm{G}}\right\|
^{2}-p_{\mathrm{AG}}\right) ,
\end{eqnarray}%
where $p_{\mathrm{AG}}:=p-\limfunc{div}[gh(\mathbf{d}\sigma
^{\flat })^{-1}\cdot \mathbf{v}_{\mathrm{A}}^{\flat }].$ HP
implies the following
equivalent equations:%
\renewcommand{\arraystretch}{1.75}%
\begin{equation}
\fbox{$%
\begin{array}{l}
\text{a. }\partial _{t}\left( \mathbf{v}_{\mathrm{G}}+\sigma \right) ^{\flat
}+\pounds _{\mathbf{v}}\left( \mathbf{v}_{\mathrm{G}}+\sigma \right) ^{\flat
}+\mathbf{d}\hspace{-0.03in}\left( p_{\mathrm{AG}}-\tfrac{1}{2}\left\| \mathbf{v}%
_{G}\right\| ^{2}-\langle \hspace{-0.02in}\langle \mathbf{v},\mathbf{v}_{%
\mathrm{G}}+\sigma \rangle \hspace{-0.02in}\rangle \right)\hspace{-0.05in}=0, \\
\text{b. }\partial _{t}\mathbf{v}_{\mathrm{G}}^{\flat
}+\mathbf{d}\left(
\mathbf{v}_{\mathrm{G}}+\sigma \right) ^{\flat }\cdot \mathbf{v}+\mathbf{d}\hspace{-0.03in}%
\left( p_{\mathrm{AG}}+\tfrac{1}{2}\left\|
\mathbf{v}_{\mathrm{G}}\right\|
^{2}\right)\hspace{-0.05in}=0, \\
\text{c. }(\partial _{t}+\nabla
_{\mathbf{v}})\mathbf{v}_{\mathrm{G}}^{\flat
}+\mathbf{dv}_{\mathrm{G}}^{\flat }\cdot \mathbf{v}_{\mathrm{A}}+\mathbf{d}%
\sigma ^{\flat }\cdot \mathbf{v}+\mathbf{d}p_{\mathrm{AG}}=0.%
\end{array}%
$}  \label{L1}
\end{equation}%
\renewcommand{\arraystretch}{1.00}%
Because of the presence of the term $\langle \hspace{-0.02in}\langle \mathbf{%
v},\mathbf{v}_{\mathrm{G}}\rangle \hspace{-0.02in}\rangle $ in
(\ref{L-L1}), in addition to the no-flow boundary condition
(\ref{BC-SW}), HP also implies the following condition:
\begin{equation}
\fbox{$\langle \hspace{-0.02in}\langle (\mathbf{d}\sigma ^{\flat
})^{-1}\cdot \mathbf{v}_{\mathrm{A}}^{\flat },n\rangle \hspace{-0.02in}%
\rangle =0\quad @\quad \partial D.$}  \label{BC-L1}
\end{equation}%
Any set selected from (\ref{L1}) (or the corresponding contravariant
counterpart through the metric) together with the volume conservation
equation (\ref{vol}), all subject to boundary conditions (\ref{BC-SW})
\textit{and }(\ref{BC-L1}), constitute the invariant form of the L1 model on
a region $D$ on the sphere. Since $\mathbf{v}_{\mathrm{G}}$ and $h$ are not
independent the L1 system has only one scalar prognostic equation; the other
two scalar equations provide the constraints to determine $\mathbf{v}_{%
\mathrm{A}}$.

The L1 model conserves geostrophic versions of the SW energy and Casimirs,
namely
\begin{equation}
\mathcal{E}_{\mathrm{G}}:=\tfrac{1}{2}\int_{D}\mathrm{d}^{2}\mathbf{x}\,%
\sqrt{m}\left( h\left\| \mathbf{v}_{\mathrm{G}}\right\| ^{2}+p^{2}/g\right)
,\quad \mathcal{C}_{\mathrm{G}}:=\int_{D}\mathrm{d}^{2}\mathbf{x}\,\sqrt{m}%
hC(q_{\mathrm{G}})
\end{equation}%
for arbitrary $C(\cdot )$, where%
\begin{equation}
q_{\mathrm{G}}h:=\frac{1}{\sqrt{m}h_{0}}\varepsilon ^{ij}\partial _{i}\left(
J\frac{\delta L_{1}}{\delta v^{j}}\right) =\frac{1}{\sqrt{m}}\varepsilon
^{ij}\partial _{i}\left( v_{\mathrm{G}j}+\sigma _{j}\right)
\end{equation}%
defines the geostrophic potential vorticity $q_{\mathrm{G}}$,
which is
materially conserved as is advected by the \textit{total} flow (i.e. $%
\mathrm{D}q_{G}/\mathrm{D}t=0$).

In geographic coordinates, the physical counterpart of, for instance, set (%
\ref{L1}b) is given by%
\renewcommand{\arraystretch}{1.25}%
\begin{equation}
\left.
\begin{array}{r}
\partial _{t}u_{\mathrm{G}}-(\xi _{\mathrm{G}}+f)v+\gamma ^{-1}\partial
_{x}B_{\mathrm{AG}}=0, \\
\partial _{t}v_{\mathrm{G}}+(\xi _{\mathrm{G}}+f)u+\partial _{y}B_{\mathrm{AG%
}}=0,%
\end{array}%
\hspace{-0.05in}\right\}  \label{L1-physical}
\end{equation}%
\renewcommand{\arraystretch}{1.00}%
where $\xi _{\mathrm{G}}:=\gamma ^{-1}\partial _{x}v_{\mathrm{G}}-\partial
_{y}u_{\mathrm{G}}-\tau u_{\mathrm{G}},\;B_{\mathrm{AG}}:=p_{\mathrm{AG}}+%
\tfrac{1}{2}(u_{\mathrm{G}}^{2}+v_{\mathrm{G}}^{2}),$ and $p_{\mathrm{AG}}$ $%
=$ $p+\gamma ^{-1}\partial _{x}\left( ghv_{\mathrm{A}}/f\right) $ $-$ $%
\partial _{y}\left( ghu_{\mathrm{A}}/f\right) $ $-$ $\tau ghu_{\mathrm{A}%
}/f. $ In vector notation set (\ref{L1-physical}) expresses as%
\begin{equation}
\partial _{t}\mathbf{u}_{\mathrm{G}}+(\xi _{\mathrm{G}}+f)\,\mathbf{\hat{z}}%
\times \mathbf{u}+\nabla B_{\mathrm{AG}}=0,
\end{equation}%
where $\xi _{\mathrm{G}}=\nabla \cdot (\mathbf{u}_{\mathrm{G}}\times \mathbf{%
\hat{z}}),$ $B_{\mathrm{AG}}=p_{\mathrm{AG}}+\tfrac{1}{2}\mathbf{u}_{\mathrm{%
G}}^{2},$ and $p_{\mathrm{AG}}=p+\nabla \cdot \left( gh\mathbf{u}_{\mathrm{A}%
}\times \mathbf{\hat{z}}/f\right) .$ Boundary condition (\ref{BC-L1}), in
turn, takes the form%
\begin{equation}
\mathbf{u}_{\mathrm{A}}\cdot \mathbf{\hat{z}}\times \mathbf{\hat{n}}=0\quad
@\quad \partial D
\end{equation}
(cf. \opencite{Ren-Shepherd-97} for a physical interpretation of
this condition). The set of diagnostic equations which determines
$\mathbf{u}_{\mathrm{A}}$
is given by%
\renewcommand{\arraystretch}{1.25}%
\begin{equation}
\left.
\begin{array}{r}
\mathsf{A}(hv_{\mathrm{A}})+\mathsf{B}(hu_{A})=F_{1}, \\
\mathsf{A}((g/f)hv_{\mathrm{A}})+(g/f)\mathsf{B}(hu_{A})=F_{2},%
\end{array}%
\hspace{-0.05in}\right\}  \label{Eq-uA}
\end{equation}%
\renewcommand{\arraystretch}{1.00}%
where the differential operators%
\begin{equation}
\mathsf{A}(\cdot ):=\nabla ^{2}(\cdot )-R^{-2}-\tau ^{2}-(f/g)q_{\mathrm{G}%
},\quad \mathsf{B}(\cdot ):=(f^{\prime }/f)\gamma ^{-1}\partial _{x}(\cdot ),
\end{equation}%
and the functions%
\begin{eqnarray}
F_{1}(h)&:\hspace{-.09in}&=-\partial _{y}\nabla \cdot (h%
\mathbf{u}_{\mathrm{G}})+(f/g)\left( hq_{\mathrm{G}}v_{\mathrm{G}}\hspace{%
-0.02in}-\gamma ^{-1}\partial _{x}B_{\mathrm{G}}\right) , \\
F_{2}(h)&:\hspace{-.09in}&=-(g/f)\gamma ^{-1}\partial
_{x}\nabla \cdot (h\mathbf{u}_{\mathrm{G}})+hq_{\mathrm{G}}u_{\mathrm{G}%
}+\partial _{y}B_{\mathrm{G}};
\end{eqnarray}%
here, $\nabla ^{2}a:=\nabla \cdot \nabla a=\gamma ^{-2}\partial
_{xx}a+\partial _{yy}a-\tau \partial _{y}a$ is the Laplacian of any scalar
function $a(\mathbf{x})$ in geographic coordinates\footnote{%
Because $\gamma ^{2}>0$ (excluding, of course, the poles) the
elliptic problem (\ref{uA}) has a unique solution on $D$ (bounded
or periodic in one or both directions) provided that
$q_{\mathrm{G}}f\geq -g(R^{-2}+\tau ^{2})$ (cf.
\opencite{Courant-Hilbert-62}), which holds for all time because
$q_{\mathrm{G}}\sim f/h$ as $\varepsilon \rightarrow 0.$}. For
completeness, from (\ref{Eq-uA}) it follows
\begin{subequations}
\label{uA}
\begin{eqnarray}
hu_{A}&=&\left( \mathsf{A}(g/f)-(g/f)\mathsf{%
BA}^{-1}\mathsf{B}\right) ^{-1}\left( F_{2}-(g/f)\mathsf{BA}%
^{-1}F_{1}\right) , \\
hv_{A}&=&\left( \mathsf{A}-\mathsf{B}(f/g)%
\mathsf{A}^{-1}(g/f)\mathsf{B}\right) ^{-1}\left( F_{1}-\mathsf{B}(f/g)%
\mathsf{A}^{-1}F_{2}\right) ,
\end{eqnarray}%
which upon substitution in the volume conservation equation
(\ref{vol}) results in a single evolution equation for the height
field. The (Cartesian) $f$-plane version of the latter was derived
by  \inlinecite{Vanneste-Bokhove-02} using a Dirac-bracket
approach. Finally, the integrals of motion of the L1 system read
\end{subequations}
\begin{equation}
\mathcal{E}_{\mathrm{G}}:=\tfrac{1}{2}\int_{D}\mathrm{d}^{2}\mathbf{x}%
\,\gamma \left( h\mathbf{u}_{\mathrm{G}}^{2}+p^{2}/g\right) ,\quad \mathcal{C%
}_{\mathrm{G}}:=\int_{D}\mathrm{d}^{2}\mathbf{x}\,\gamma hC(q_{\mathrm{G}%
}),
\end{equation}%
where $q_{\mathrm{G}}=(\xi _{\mathrm{G}}+f)/h$; as before if $D$
and $H$ are zonally symmetric then
\begin{equation}
\mathcal{M}_{\mathrm{G}}:=\int_{D}\mathrm{d}^{2}\mathbf{x}\,\gamma h\left[
\gamma u_{\mathrm{G}}-\Omega R\left( \cos \vartheta _{0}-\gamma \cos
\vartheta \right) \right]
\end{equation}%
is also conserved.

Other decompositions, appart than (\ref{v-split}), as well as
other balance relationships, different than (\ref{vG}), are possible \cite%
{Allen-Holm-96,Allen-Holm-Newberger-02}. This freedom is what
allows for the existence of approximate models which can be
potentially more accurate than the L1 model.

\section{Concluding Remarks\label{Conclusion}}

The scaling%
\begin{equation}
\{\mathbf{u},\partial _{t},y/R,h-H(y),H^{\prime }\}=O(\varepsilon )
\end{equation}%
implies, at $O(\varepsilon ^{2})$, the classical QG equation (cf.
\opencite{Pedlosky-87})
\begin{subequations}
\label{QG}
\begin{equation}
\left( \partial _{t}+\partial _{x}\psi \partial _{y}-\partial _{y}\psi
\partial _{x}\right) q_{\mathrm{QG}}=0,
\end{equation}
where
\begin{equation}
q_{\mathrm{QG}}:=\left[
\partial _{xx}+\partial _{yy}-f_{0}^{2}/(gH_{0})\right] \psi +\left( \beta
+\beta _{\mathrm{T}}\right) y.
\end{equation}
\end{subequations}
Here, $H(y)=H_{0}(1-\beta _{\mathrm{T}}y/f_{0})$, with $H_{0}=\mathrm{%
\limfunc{const}.}$, and $\psi :=g[h-H(y)]/f_{0}$ is the
geostrophic streamfunction, i.e. $\mathbf{u}=(-\partial _{y}\psi
,\partial _{x}\psi )+O(\varepsilon ^{2})$. (More complicated
topographies can of course be considered.)
Notice the absence of geometric coefficients in (\ref%
{QG}). Those terms, which \textit{do} appear in the corresponding
(diagnostic) momentum and volume conservation equations, have (fortuitously)%
\textit{\ }cancelled out in the process of constructing the (prognostic)
potential vorticity equation (\ref{QG}) [Pedlosky 1987; R97]\nocite%
{Pedlosky-87,Ripa-JPO-97b}. Consequently---and remarkably---QG flows develop
\textit{as if} the geometry were Cartesian, ``feeling'' the latitudinal
variation of the Coriolis parameter as the \textit{only} effect of the
Earth's sphericity.

The L1 model shares a series of differences and similarities with
the above QG model. Although both models are derivable from HP,
the QG model's action is not seen to derive from approximations
performed in SW's action. As QG motions, those governed by the L1
model are not allowed at the equator, i.e. where $f$ vanishes. In
addition to Rossby waves, the linear waves of the L1 model include
(a form of) Kelvin waves, which are not supported by the QG model.
Unlike QG motions, L1 motions are restricted neither to
$O(\varepsilon)$ meridional excursions nor to $O(\varepsilon)$
displacements of the free surface from the position of equilibrium
at rest, nor to the presence of $O(\varepsilon)$ topographic
variations. In a reduced-gravity setting, the equations for both
SW and L1 models have the same structure as those presented here,
except that in that case $g$ must be identified with the buoyancy
jump at the interface between
the active and the quiescent (infinitely deep) bottom layer, and $H(\mathbf{x%
})$ must be understood as the nonuniform thickness of the active layer at
rest, including the possibility of a nonspherical rigid surface.
Consequently, in contrast to the QG model, the L1 model is able to describe
the dynamics of frontal structures.

The integrals of motion of the L1 system, in geographic coordinates, expand
in inverse powers of the radius of the spherical Earth $R$ as%
\begin{eqnarray}
\mathcal{E}_{\mathrm{G}}\hspace{-.075in}&=&\hspace{-.075in} \left\langle \hspace{-0.025in}\left( \frac{1}{2}-\frac{\beta y}{%
f_{0}}\right) \frac{(\partial _{x}p)^{2}+(\partial _{y}p)^{2}}{f_{0}^{2}}+%
\frac{\tau _{0}y}{f_{0}^{2}}(\partial _{x}p)^{2}+\frac{p^{2}}{2gh}\hspace{-0.01in}\right\rangle  \\
\hspace{-1in}\mathcal{M}_{\mathrm{G}}\hspace{-.075in}&=&\hspace{-.075in}
\left\langle \hspace{-0.025in}\left( \frac{2\tau
_{0}f_{0}+\beta }{f_{0}}y\hspace{-0.025in}-\hspace{-0.025in}1\right) \frac{%
\partial _{y}p}{f_{0}}\hspace{-0.025in}-\hspace{-0.025in}\left( 1\hspace{%
-0.025in}-\hspace{-0.025in}\tau _{0}y\right) f_{0}y\hspace{-0.025in}-\hspace{%
-0.025in}\frac{1}{2}\beta \left(
1\hspace{-0.025in}-\hspace{-0.025in}R^{2}\tau
_{0}^{2}\right) y^{2}\hspace{-0.025in}\right\rangle  \\
\mathcal{C}_{\mathrm{G}}\hspace{-.075in}&=&\hspace{-.075in}
\left\langle (1-\tau _{0}y)C(q_{0})+q_{1}C^{\prime
}(q_{0})\right\rangle
\end{eqnarray}%
$+$ $O(R^{-2}).$ Here, $\langle\cdot\rangle:=\int_{D}\mathrm{d}^{2}\mathbf{x}%
\,h\,(\cdot ),$ and
\begin{eqnarray}
q_{0}h\hspace{-.05in}&:\hspace{-.09in}&=\frac{\partial _{xx}p+\partial _{yy}p}{f_{0}}+f_{0}, \\
q_{1}h\hspace{-.05in}&:\hspace{-.09in}&=\frac{(2\tau
_{0}f_{0}-\beta )\partial _{xx}p-\beta
\partial _{yy}p}{f_{0}^{2}}y+\frac{\tau _{0}f_{0}-\beta
}{f_{0}^{2}}\partial _{y}p+\beta y.
\end{eqnarray}%
Clearly, a consistent (not necessarily the optimal, though)
geometric approximation for L1 dynamics, which is first-order accurate in $R$%
, is given by the non-Cartesian Ripa ``plane'' and \textit{not} by
the standard $\beta $ plane (recall that it has $\tau _{0}=0$).
The latter is the one included in the original derivation of the
L1 model. An important contribution of the present work to the
above list of differences and similarities between the L1 and QG
models is thus the sensitivity of the former to the differences
between geographic and geodesic coordinates. This result confirms
\citeauthor{Ripa-JPO-00b}'s \shortcite{Ripa-JPO-00b} in the sense
that Earth's curvature effects increase in importance as the
motions deviate from strictly geostrophic (divergence-free)
motions. The thorough evaluation of these effects, apart from
checking that the equations have the right conservation laws, is a
subject for futher research. The latter should involve direct
numerical simulations in which predictions of the L1 model on the
$\beta $-plane and the sphere (or the Ripa ``plane'') are
compared.

Finally, \inlinecite{Ripa-JFM-83} showed that steady SW flows on
the sphere posses a formal stability theorem. The latter involving
an Arnold-like first theorem for the stability of QG flows, and a
condition for the flow to be ``subsonic'' in the sense that the
(geostrophic) basic flow must be everywhere slower than the
slowest gravity-wave of the system. \inlinecite{Ren-Shepherd-97}
showed, in turn, that steady L1 flows on the $\beta $ plane posses
a Ripa-like formal stability theorem, as well as a nonlinear (or
Lyapunov) stability theorem in which the ``subsonic'' condition of
Ripa's theorem is replaced by a condition that the flow be
cyclonic along the lateral boundaries. The latter was shown by
\inlinecite{Ren-Shepherd-97} to have an interpretation involving
coastal Kelvin waves, which are not included in the QG model.
Whether or not L1 flows on the sphere (or the Ripa ``plane'')
enjoy similar stability properties is an issue that needs more
investigation.

\acknowledgements

Part of this work was carried out while I was an ScD student at
CICESE (Mexico) under the supervision of Pedro Ripa. His untimely
death will not prevent me from finding in him a source of
inspiration. I have benefited from fruitful conversations with
Alejandro Par\'{e}s, Julio Sheinbaum, M. Josefina Olascoaga, and
Oscar U. Velasco-Fuentes. Corrections of the manuscript by M.
Josefina Olascoaga and Michael G. Brown are sincerely appreciated.
The comments of an anonymous reviewer lead to improvements in the
paper. My work was partly supported by CICESE, CONACyT (Mexico),
and NSF (USA).

\def\thesection{\Alph{section}}
\setcounter{section}{0}
\def\theequation{\Alph{section}.\arabic{equation}}
\setcounter{equation}{0}

\section{Useful Relations\label{Relations}}

The following relationships can be shown to hold:
\begin{subequations}
\label{Js}
\begin{gather}
J\mathfrak{J}=1\Longleftrightarrow J_{\mathfrak{i}}^{i}\mathfrak{J}_{j}^{%
\mathfrak{i}}=\delta _{j}^{i}, \\
(\limfunc{adj}J)_{i}^{\mathfrak{i}}=J\mathfrak{J}_{i}^{\mathfrak{i}%
}\Longleftrightarrow J=J_{\mathfrak{i}}^{i}(\limfunc{adj}J)_{i}^{\mathfrak{i}%
}\Longleftrightarrow \partial J/\partial J_{\mathfrak{i}}^{i}=J\mathfrak{J}%
_{i}^{\mathfrak{i}}, \\
(\limfunc{adj}\mathfrak{J})_{\mathfrak{i}}^{i}=\mathfrak{J}J_{\mathfrak{i}%
}^{i}\Longleftrightarrow \mathfrak{J}=\mathfrak{J}_{i}^{\mathfrak{i}}(%
\limfunc{adj}\mathfrak{J})_{\mathfrak{i}}^{i}\Longleftrightarrow \partial
\mathfrak{J}/\partial \mathfrak{J}_{i}^{\mathfrak{i}}=\mathfrak{J}J_{%
\mathfrak{i}}^{i}, \\
\partial _{\mathfrak{i}}\left( J\mathfrak{J}_{i}^{\mathfrak{i}}\right)
=0=\partial _{i}\left( \mathfrak{J}J_{\mathfrak{i}}^{i}\right) , \\
\delta J=J\partial _{i}\delta x^{i},\;\delta \mathfrak{J}=\mathfrak{J}%
\partial _{\mathfrak{i}}\delta \mathfrak{x}^{\mathfrak{i}}, \\
\partial _{\mathfrak{i}}=J_{\mathfrak{i}}^{i}\partial _{i},\;\partial _{i}=%
\mathfrak{J}_{i}^{\mathfrak{i}}\partial _{\mathfrak{i}}.
\end{gather}%
In deriving (\ref{Js}e) the following properties of the determinants were
very helpful
\end{subequations}
\begin{equation}
\dfrac{\partial (a,b)}{\partial (\mathfrak{x}^{1},\mathfrak{x}^{2})}=\dfrac{%
\partial (a,b)}{\partial (x^{1},x^{2})}\dfrac{\partial (x^{1},x^{2})}{%
\partial (\mathfrak{x}^{1},\mathfrak{x}^{2})},\quad \dfrac{\partial (a,x^{2})%
}{\partial (x^{1},x^{2})}=\frac{\partial a}{\partial x^{1}}
\end{equation}%
for all scalar functions $a,b(\mathbf{x}).$

In addition, it can be shown that:
\begin{subequations}
\label{relations2}
\begin{gather}
\mathfrak{\dot{x}}^{\mathfrak{i}}=\partial _{t}\mathfrak{x}^{\mathfrak{i}%
}+v^{i}\mathfrak{J}_{i}^{\mathfrak{i}}=0\Longrightarrow v^{i}=-\mathfrak{J}_{%
\mathfrak{i}}^{i}\partial _{t}\mathfrak{x}^{\mathfrak{i}}, \\
\partial _{\mathfrak{i}}v^{i}=J_{\mathfrak{i}}^{j}\partial
_{j}v^{i}\Longrightarrow \dot{J}_{\mathfrak{i}}^{i}=J_{\mathfrak{i}%
}^{j}\partial _{j}v^{i}, \\
\delta v^{i}=-J_{\mathfrak{i}}^{i}(\partial _{t}+v^{j}\partial _{j})\delta
\mathfrak{x}^{\mathfrak{i}},\quad \delta \mathfrak{J}=\mathfrak{J}J_{%
\mathfrak{i}}^{i}\partial _{i}\delta \mathfrak{x}^{\mathfrak{i}}, \\
\delta J=\mathfrak{J}\partial _{i}\delta x^{i}\Longrightarrow \partial _{t}%
\mathfrak{J}+\partial _{i}\left( \mathfrak{J}v^{i}\right) =0.
\end{gather}

\setcounter{equation}{0}

\section{Alternative HPs}

\subsection{Eulerian Coordinates}

The standard approach for fields $\mathfrak{x}(\mathbf{x},t)$
consists in considering an action functional of the form
\end{subequations}
\begin{equation}
\mathcal{S}[\mathfrak{x}]:=\int_{t_{0}}^{t_{1}}\mathrm{d}t\,L[\mathfrak{x}%
,\partial _{t}\mathfrak{x}]=\int_{t_{0}}^{t_{1}}\mathrm{d}t\int_{D}\mathrm{d}%
^{2}\mathbf{x\,}l(\mathfrak{x},\partial _{t}\mathfrak{x},\mathfrak{J}_{i}^{%
\mathfrak{i}};\mathbf{x},t),  \tag{B.1}
\end{equation}%
which, after invoking HP results in the familiar Euler--Lagrange equations%
\begin{equation}
\partial _{t}\frac{\delta L}{\delta \mathfrak{x}_{,t}^{\mathfrak{i}}}-\frac{%
\delta L}{\delta \mathfrak{x}^{\mathfrak{i}}}=\partial _{t}\frac{\partial l}{%
\partial \mathfrak{x}_{,t}^{\mathfrak{i}}}+\partial _{i}\frac{\partial l}{%
\partial \mathfrak{J}_{i}^{\mathfrak{i}}}-\partial _{\mathfrak{i}}l=0
\tag{B.2}  \label{standard}
\end{equation}%
(plus boundary conditions). According to $\partial _{i}(\partial
l/\partial
\mathfrak{J}_{i}^{\mathfrak{i}})=\partial _{i}(\mathfrak{J}J_{\mathfrak{i}%
}^{i}\partial l/\partial \mathfrak{J)}=\mathfrak{J}J_{\mathfrak{i}%
}^{i}\partial _{i}(\partial l/\partial \mathfrak{J)}$ and (\ref{relations2}%
a), equations (\ref{standard}) transform into%
\begin{equation}
\partial _{t}\left( J_{\mathfrak{i}}^{i}\frac{\partial l}{\partial v^{i}}%
\right) -\mathfrak{J}J_{\mathfrak{i}}^{i}\partial _{i}\frac{\partial l}{%
\partial \mathfrak{J}}+\partial _{\mathfrak{i}}l=0.  \tag{B.3}
\end{equation}%
The latter can be shown to be equivalent to (\ref{EulerLagrange})
only in the particular case $\partial _{\mathfrak{i}}l\equiv 0.$

\subsection{Lagrangian Coordinates}

In the variational approach for fields
$\mathbf{x}(\mathfrak{x},t)$ the
action is of the form%
\begin{equation}
\mathcal{S}[\mathbf{x}]:=\int_{t_{0}}^{t_{1}}\mathrm{d}t\,L[\mathbf{x}%
,\partial _{t}\mathbf{x}]=\int_{t_{0}}^{t_{1}}\mathrm{d}t\int_{\mathfrak{D}}%
\mathrm{d}^{2}\mathfrak{x\,}\mathbf{\,}l(\mathbf{x},\partial _{t}\mathbf{x}%
,J_{\mathfrak{i}}^{i};\mathfrak{x},t).  \tag{B.4}
\end{equation}%
Upon variations of particle paths at fixed Lagrangian labels and
time, the
following Euler--Lagrange equations result from HP:%
\begin{equation}
\partial _{t}\frac{\delta L}{\delta x_{,t}^{i}}-\frac{\delta L}{\delta x^{i}}%
=\partial _{t}\frac{\partial l}{\partial x_{,t}^{i}}+\partial _{\mathfrak{i}}%
\frac{\partial l}{\partial J_{\mathfrak{i}}^{i}}-\partial _{i}l=0
\tag{B.5}
\end{equation}%
(plus appropriate boundary conditions). This is but the
infinite-dimensional analogue of the particle's HP. For instance,
S83's derivation of the SW and L1 systems in the Cartesian
coordinates of the $\beta $ plane and
\citeauthor{vanderToorn-97}'s \shortcite{vanderToorn-97}
derivation of the SW equations on the sphere are based on this HP.
One disadvantage of this variational approach, however, is that
the resulting equations are in Lagrangian coordinates, which
requires application of the inverse map $\varphi ^{-1}$
(\ref{inverse}) to transform back to Eulerian coordinates.

\bibliographystyle{klunamed}

\end{document}